# Laser Controlled Spin Dynamics of Ferromagnetic Thin Film from Femtosecond to Nanosecond Timescale

Sucheta Mondal and Anjan Barman[*]

Department of Condensed Matter Physics and Material Sciences, S. N. Bose National Centre for Basic Sciences, Block JD, Sector III, Salt Lake, Kolkata 700 106, India.

[*]abarman@bose.res.in

Laser induced modulation of the magnetization dynamics occurring over various time-scales have been unified here for a $Ni_{80}Fe_{20}$ thin film excited by amplified femtosecond laser pulses. The weak correlation between demagnetization time and pump fluence with substantial enhancement in remagnetization time is demonstrated using three-temperature model considering the temperatures of electron, spin and lattice. The picosecond magnetization dynamics is modeled using the Landau-Lifshitz-Gilbert equation. With increasing pump fluence the Gilbert damping parameter shows significant enhancement from its intrinsic value due to increment in the ratio of electronic temperature to Curie temperature within very short time scale. The precessional frequency experiences noticeable red shift with increasing pump fluence. The changes in the local magnetic properties due to accumulation and dissipation of thermal energy within the probed volume are described by the evolution of temporal chirp parameter in a comprehensive manner. A unification of ultrafast magnetic processes and its control over broad timescale would enable the integration of various magnetic processes in a single device and use one effect to control another.

# I. INTRODUCTION

Recent development in magnetic storage [1] and memory [2] devices heavily relies upon increasing switching speed and coherent switching of magnetic states in ferromagnetic thin films and patterned structures. Operating speeds of information storage devices have progressed into the sub-gigahertz regime and controlled switching in individual layers of magnetic multilayers and heterostructures has become important. The relaxation processes involved in magnetization dynamics set natural limits for these switching times and data transfer rates. In the context of precessional magnetization dynamics the natural relaxation rate against the small perturbation is expressed as Gilbert damping ($\alpha$) according to the Landau-Liftshiz-Gilbert (LLG) equation [3, 4]. This is analogous to viscous damping of the mechanical frictional torque and leads to the direct dissipation of energy from the uniform precessional mode to thermal bath in case of zero wave-vector excitation. Gilbert damping originates from spin-orbit coupling and depends on the coupling strength and $d$-band width of the $3d$ ferromagnet [5]. The damping can be varied by various intrinsic and extrinsic mechanisms including phonon drag [6], Eddy current [7], doping [8] or capping [9] with other material, injection of spin current [10], magnon-magnon scattering [11] and controlling temperature of the system [12]. Intrinsic and extrinsic nature of Gilbert damping were primarily studied by using ferromagnetic resonance (FMR) technique. When the magnetization is aligned with either in-plane or out-of-plane applied magnetic field, the linewidth is proportional to the frequency with a slope determined by damping coefficient. This is the homogeneous or intrinsic contribution to the FMR linewidth. However, experiments show an additional frequency-independent contribution to the linewidth corresponds to inhomogeneous line broadening [13, 14]. However, state-of-the-art technique based on pump-probe geometry has been developed and rigorously exploited for measuring ultrafast magnetization dynamics of ferromagnetic thin films during last few decades [15, 16]. Using time-resolved magneto-optical Kerr effect (TR-MOKE) technique one can directly address the processes which are responsible for the excitation and relaxation of a magnetic system on their characteristic time scales [17-19]. Generally during the pump-probe measurements pump fluence is kept low to avoid nonlinear effects and sample surface degradation. Some recent experiments reveal that nonlinear spin waves play a crucial role in high power thin film precessional dynamics by introducing spin-wave instability [20] similar to FMR experiments by application of high rf power [21]. The coercivity and anisotropy of the ferromagnetic thin films can also be lowered by pump fluence, which may have potential application in heat assisted magnetic recording (HAMR) [22]. Recent reports reveal that damping coefficient can be increased or decreased noticeably in the higher excitation regime due to opening of further energy dissipation channels beyond a threshold pump power [23-25]. Not only relaxation parameters but also frequency shift due to enhancement in pump power has been observed [20]. However, the experimental evidence for large modulation of Gilbert damping along with frequency shift and temporal chirping of the uniform precessional motion is absent in the literature. This investigation demands suitable choice of material, and here we have chosen Permalloy ($Ni_{80}Fe_{20}$ or Py here on) because of its high permeability, negligible magneto-crystalline anisotropy, very low coercivity, large anisotropic magnetoresistance with reasonably low damping. Also, due to its negligible magnetostriction Py is less sensitive to strain and stress exerted during the thermal treatment in HAMR [22].

In this article, we have used femto-second amplified laser pulses for excitation and detection of ultrafast magnetization dynamics in a Py thin film. Pump fluence dependent ultrafast demagnetization is investigated along with fast and slow remagnetization. Our comprehensive study of the picosecond dynamics reveals transient nature of enhanced Gilbert damping in presence of high pump fluence. Also, the time-varying precession is subjected to temporal chirping which occurs due to enhancement of temperature of the probed volume within a very short time scale being followed by successive heat dissipation. This fluence dependent modulation of magnetization dynamics will undoubtedly found suitable application in spintronic and magnonic devices.

## II. SAMPLE PREPARATION AND CHARACTERIZATION

20-nm-thick Permalloy ($Ni_{80}Fe_{20}$, Py hereafter) film was deposited by using electron-beam evaporation technique (SVT Associates, model: Smart Nano Tool AVD-E01) (base pressure = 3 × $10^{-8}$ Torr, deposition rate = 0.2 Å/S) on 8 × 8 $mm^2$ silicon (001) wafer coated with 300-nm-thick $SiO_2$. Subsequently, 5-nm-thick $SiO_2$ is deposited over the $Ni_{80}Fe_{20}$ using rf sputter-deposition technique (base pressure = 4.5 × $10^{-7}$ Torr, Ar pressure = 0.5 mTorr, deposition rate = 0.2 Å/S, rf power = 60 W). This capping layer protects the surface from environmental degradation, oxidation and laser ablation during the pump-probe experiment using femtosecond laser pulses. From the vibrating sample magnetometry (VSM) we have obtained the saturation magnetization ($M_s$) and Curie temperature ($T_c$) to be 850 emu/cc and 863 K, respectively [26].

To study the ultrafast magnetization dynamics of this sample, we have used a custom-built time resolved magneto optical Kerr effect (TRMOKE) magnetometer based on optical pump-probe technique as shown in Fig. 1 (a). Here, the second harmonic (λ = 400 nm, repetition rate = 1 kHz, pulse width > 40 fs) of amplified femtosecond laser pulse generated from a regenerative amplifier system (Libra, Coherent) is used to excite the dynamics while the fundamental laser pulse (λ = 800 nm, repetition rate = 1 kHz, pulse width ≈ 40 fs) is used as probe to detect the time-resolved polar Kerr signal from the sample. The temporal resolution of the measurement is limited by the cross-correlation between the pump and probe pulses (≈120 fs). The probe beam having diameter of about 100 μm is normally incident on the sample whereas the pump beam is kept slightly defocused (spot size is about 300 μm) and is obliquely (≈ 30° with normal to the sample plane) incident on the sample maintaining an excellent spatial overlap with the probe spot. Time-resolved Kerr signal is collected from the uniformly excited part of the sample and slight misalignment during the course of the experiment does not affect the pump-probe signal significantly. A large magnetic field of 3.5 kOe is first applied at a small angle of about 10° to the sample plane to saturate its magnetization. This is followed by reduction of the magnetic field to the bias field value (*H* = in-plane component of the bias field), which ensures that the magnetization remains saturated along the bias field direction. The tilt of magnetization from the sample plane ensures a finite demagnetizing field along the direction of the pump pulse, which is further modified by the pump pulse to induce a precessional dynamics within the sample [17]. In our experiment a 2-ns time window has been used, which gave a damped uniform precession of magnetization. The pump beam is chopped at 373 Hz frequency and the dynamic signal in the probe pulse is detected by using a lock-in amplifier in a phase sensitive manner. Simultaneous time-resolved reflectivity and Kerr rotation data were measured and no significant breakthrough of one into another has been found [26]. The probe fluence is kept constant at 2 mJ/cm$^2$ during the measurement to avoid additional contribution to the modulation of spin dynamics via laser heating. Pump fluence (*F*) was varied from 10 to 55 mJ/cm$^2$ to study the fluence dependent modulation in magnetization dynamics. All the experiments were performed under ambient condition and room temperature.

### III. RESULTS AND DISCUSSIONS

### A. Laser induced ultrafast demagnetization

When a femtosecond laser pulse interacts with a ferromagnetic thin film in its saturation condition, the magnetization of the system is partially or fully lost within hundreds of femtosecond as measured by the time-resolved Kerr rotation or ellipticity. This is known as ultrafast demagnetization of the ferromagnet and was first observed by Beaurepire et al. in 1996 [27]. This is generally followed by a fast recovery of the magnetization within sub-picosecond to few picoseconds and a slower recovery within tens to hundreds of picoseconds, known as the fast and slow remagnetization. In many cases the slower recovery is accompanied by a coherent magnetization precession and damping [17]. In our pump-probe experiment, the sample magnetization is maintained in the saturated state by application of a magnetic field *H* = 2.4 kOe before zero delay. Right after the zero-delay and the interaction of the pump pulse with the electrons in the ferromagnetic metal, ultrafast demagnetization takes place. The local magnetization is immediately quenched within first few hundreds of fs followed by a subsequent fast remagnetization in next few ps [27]. Figure 1(b) shows ultrafast demagnetization obtained for different pump fluences. Several models have been proposed over two decades to explain the ultrafast demagnetization [16, 28-31]. Out of those a phenomenological thermodynamic model, called three temperature model [27, 32, 33] has been most widely used, where the dynamics of these spin fluctuations can be describes as:

$$\Delta M = \theta(t)(-A_1 + \frac{A_2\tau_{el-lat} - A_1\tau_{el-sp}}{\tau_{el-lat} - \tau_{el-sp}}e^{-(\frac{t}{\tau_{el-sp}})} + \frac{A_1 - A_2}{\tau_{el-lat} - \tau_{el-sp}}\tau_{el-lat}e^{-(\frac{t}{\tau_{el-lat}})})M_0 \otimes \Gamma(t) \quad (1).$$

This is an approximated form based on the assumption that the electron temperature rises instantaneously upon laser excitation and can be applied to fit time-resolved Kerr rotation data taken within few picoseconds timescale. The whole system is divided into three subsystems:

electron, spin and lattice system. On laser excitation the hot electrons are created above Fermi level. Then during energy rebalancing between the subsystems, quenched magnetization relaxes back to the initial state. The two exponential functions in the above equation mirror the demagnetization given by demagnetization time ($\tau_{el-sp}$) for energy transfer between electron-spin and the decay of electron temperature ($\tau_{el-lat}$) owing to the transfer of energy to the lattice. In addition to these characteristics time constants, the spin-lattice relaxation time also can be extracted by including another exponential term in the above equation if the spin specific heat is taken into account [34]. θ is the Heaviside step function and Γ(t) stands for the Gaussian function to be convoluted with the laser pulse envelope determining the temporal resolution (showing the cross correlation between the probe and pump pulse). The constant, $A_1$ indicates the ratio between amount of magnetization after equilibrium between electrons, spins, and lattice is restored and the initial magnetization. $A_2$ is proportional to the initial electronic temperature rise. We have plotted A1 and A2, normalized with their values at the highest fluence, as a function of pump fluence in Fig. 3S of the supplemental material which shows that magnitude of both parameters increases with laser fluence [26]. We have observed that with increasing fluence the demagnetization time has been negligibly varied within a range of 250±40 fs. The weak or no correlation between the pump fluence and the demagnetization rate describes the intrinsic nature of the spin scattering, governed by various mechanisms including Elliott-Yafet mechanism [35]. Another important observation here is that the delay of demagnetization processes which is the time delay between pump pulse (full width at half maxima, FWHM ≈ 130±20 fs) and starting point of the ultrafast demagnetization, becomes shorter due to increase in pump fluences. A plausible explanation for this is the dependence of delay of demagnetization on the electron-thermalization time which is eventually proportional to electron density or pump fluences [36]. On the other hand, fast remagnetization time has been found to be increased noticeably from 0.40 ± 0.05 ps to 0.80 ± 0.05 ps within the experimental fluence range of 10-55 mJ/cm$^2$. The larger is the pump fluence, the higher is the electron temperature or further the spin temperature. Therefore, it is reasonable that magnetization recovery time increases with the pump fluence.

**B. Pump fluence dependent modulation in Gilbert damping**
Figure 1 (c) shows the representative Kerr rotation data for $F$ = 25 mJ/cm$^2$ consisting of three temporal regions, *i.e.* ultrafast demagnetization, fast remagnetization and slow remagnetization superposed with damped precession within the time window of 2 ns. We process the magnetization precession part after subtracting a bi-exponential background to estimate the damping and its modulation. The slower remagnetization is mainly due to heat diffusion from the lattice to the substrate and surrounding. Within our experimental fluence range the slow remagnetization time has increased from ≈0.4 ns to ≈1.0 ns. The precessional dynamics is described by phenomenological Landau-Lifshitz-Gilbert (LLG) equation,

$$\frac{d\vec{M}}{dt} = -\gamma\, \vec{M} \times \vec{H}_{eff} + \frac{\alpha}{M_s} \vec{M} \times \frac{d\vec{M}}{dt} \qquad (2)$$

where $\gamma$ is the gyromagnetic ratio, $M$ is magnetization, $\alpha$ is Gilbert damping constant and $H_{eff}$ is the effective magnetic field consisting of several field components. The variation of precessional frequency with the angle between sample plane and bias magnetic field direction is plotted in Fig. 1 (d), which reveals that there is no uniaxial anisotropy present in this sample.

The energy deposited by the pump pulse, in terms of heat within the probed volume, plays a very crucial role in modification of local magnetic properties, i.e. magnetic moment, anisotropy, coercivity, magnetic susceptibility, etc. With increasing fluence the precessional frequency experienced a red shift [20, 25]. Thus, at the onset of the precessional dynamics (about 10 ps from zero delay), for relatively high fluence, the initial frequency ($f_i$) will be smaller than its intrinsic value (in absence of any significant heat dissipation). As time progresses and the sample magnetization gradually attains its equilibrium value, the precessional frequency continuously changes, causing a temporal chirping of the damped oscillatory Kerr signal. The frequency shift can be estimated from the amount of temporal chirping [37]. Figure 2 (a) shows the background subtracted time-resolved Kerr rotation data (precessional part) for different pump fluences fitted with a damped sinusoidal function with added temporal chirping,

$$\theta_k = A e^{-t/\tau} \sin(2\pi(f_i + bt)t + \Phi)$$

where $A$, $\tau$, $f_i$, $b$ and $\Phi$ are the amplitude of the magnetization precession, the relaxation time, the initial precessional frequency, chirp parameter and initial phase, respectively. At this point, we are unsure of the exact nature of the damping, *i.e.* it may consist of both intrinsic and extrinsic mechanisms and hence we term it as effective damping parameter ($\alpha_{eff}$) which can be extracted using the following formula [38],

$$\alpha_{eff} = \frac{1}{\gamma \tau (H + \frac{4\pi M_{eff}}{2})} \tag{3}$$

$\gamma$ = 1.83 ×10$^7$ Hz/Oe for Py and $M_{eff}$ is the effective magnetization including pump-induced changes at $H$ = 2.4 kOe. This formula is exploited to extract effective damping parameter precisely in the moderate bias field regime. The variation of relaxation time and effective damping are plotted with pump fluence in Fig. 2 (b) and (c). Here, $\tau$ decreases with fluence while damping increases significantly with respect to its intrinsic value within this fluence range. We have repeated the experiment for two different field values (2.4 and 1.8 kOe). The slope of fluence dependent damping remains unaltered for both the field values. We have also observed increase in relative amplitudes of precession with pump fluence as shown in the inset of Fig. 2 (c). To verify the transient nature of damping we have performed another set of experiment where the probed area is exposed to different pump fluences ($F_i$) for several minutes. After the irradiation, the precessioanl dynamics is measured from that area with fixed probe and pump fluences 2 and 10 mJ/cm$^2$, respectively. We found that damping remains almost constant for all the measurements (as shown in Fig. 2 (d)). These results demonstrate that the enhancement of damping is transient and only exists in the presence of high pump fluence but dropped to its original value when the pump laser was set to initial fluence.

The bias field dependence of precessional dynamics at four different pump fluences is studied to gain more insight about the origin of fluence dependent damping. First, we plotted the average frequency ($f_{FFT}$) with bias field which is obtained from the fast Fourier transformation (FFT) of the precessional data in Fig. 3 (a). The experimental data points are fitted with the Kittel formula,

$$f_{FFT} = \frac{\gamma}{2\pi}\sqrt{H(H + 4\pi M_{eff})} \qquad (4)$$

$M_{eff}$ is the effective magnetization of the sample. Figure 3 (b) shows that effective magnetization does not vary much within the applied fluence range. So, we infer that with increasing fluence there is no induced anisotropy developed in the system which can modify the effective damping up to this extent [23]. The variation of relaxation time with bias field for four different pump fluences are plotted in Fig. 3 (c). Relaxation time is increased with decreasing field for each case but for the higher fluence regime, those values seem to be fluctuating. This dependence of τ on field was fitted with equation 3 to extract damping coefficient at different fluence values. We have further plotted the damping coefficient as a function of precession frequency ($f_{FFT}$) [see supplemental material, Fig. 4S] [26], which shows an invariance of $α_{eff}$ with $f_{FFT}$. From that we can infer that the damping coefficient in our sample within the experimental field and fluence regime are intrinsic in nature and hence, we may now term it as the intrinsic damping coefficient $α_0$. The extrinsic contributions to damping mainly come from magnetic anisotropy field, two-magnon scattering, multimodal dephasing for excitation of several spin-wave modes, etc, which are negligible in our present case.

Figure 3(d) shows the variation of $α_0$ with pump fluence, which shows that even the intrinsic damping is significantly increasing with pump fluence [20, 39]. For generation of perpendicular standing spin-wave modes the film needs to be thick enough. Though the film thickness is 20 nm here, but within the applied bias field range we have not found any other magnetic mode appearing with the uniform Kittel mode within the frequency window of our interest (as shown in Fig. 5S of supplemental material) [26]. Also, for 20-nm-thick Py film, the effect of eddy current will be negligible [40]. The overlap between spatial profile of focused probe and pump laser spot may lead to the generation of magnons that propagate away from the region that is being probed. Generally, enhancement of nonlocal damping by spin-wave emission becomes significant when the excitation area is less than 1 μm. Recently J. Wu *et al.* showed that propagation of magnetostatic spin waves could be significant even for probed regions of tens of microns in size [41]. Also, by generating spin-wave trap in the pump-probe experiment modification of precessional frequency in ferromagnetic thin film due to accumulation and dissipation of thermal energy within the probed volume has been reported [42]. During our experiment the overlap between probe and pump spot is maintained carefully and Kerr signal is collected from the uniformly excited part of the sample so that slight misalignment during the course of experiment does not introduce any nonlocal effects. We will now substantiate our results with some theoretical arguments which involve the calculation of electronic temperature rise in the system due to application of higher pump fluence. The electronic temperature ($T_e$) is related to absorbed laser energy per unit volume ($E_a$) according to the following equation [43],

$$E_a = \xi(T_e^2 - T_0^2)/2 \tag{5}$$

where, $\xi$ is the electronic specific heat of the system and $T_0$ is the initial electronic temperature (room temperature here). First, we have estimated $E_a$ according to the optical parameters of the sample by using the following equation,

$$E_a = [(1 - e^{-\frac{d}{\Psi}})F(1-R)/d] \tag{6}$$

where, $d$ is sample thickness, $\Psi$ is optical penetration depth (~17 nm for 400-nm pump laser in 20-nm-thick Py film), $R$ is the reflectivity of the sample (0.5 measured for the Py film) and $F$ is applied pump fluence. By solving equations (5) and (6) we have observed that $T_e$ increases from ≈1800 to 4500 K within our experimental fluence range of 10 to 55 mJ/cm$^2$. Decay time of the electron temperature and other relevant parameters (i.e. $E_a$, $T_e$ at various fluences) are described in the supplemental material [26]. The sample remains in its magnetized state even if the electronic temperature exceeds the Curie temperature $T_c$. Importantly, ratio of the system temperature, $T$ (as decay of electronic temperature is strongly correlated with rise of lattice temperature) to $T_c$ is affecting the magnetization relaxation time which fundamentally depends on susceptibility. Accordingly damping should be proportional to susceptibility which is strongly temperature dependent [40]. Various procedures for exciting precessional dynamics in ferromagnets show the different mechanisms to be responsible for exploration of different energy dissipation channels. The spin-phonon interaction mechanism, which historically has been thought to be the main contribution to magnetization damping, is important for picosecond-nanosecond applications at high temperatures such as spin caloritronics. But for laser-induced magnetization dynamics, where spin-flips occur mainly due to electron scattering, quantum Landau-Lifshitz-Bloch equation is sometimes exploited to explain the temperature dependence of damping by considering a simple spin-electron interaction as a source for magnetic relaxation [44]. This approach suggests that increasing ratio between system temperature and Curie temperature induces electron-impurity led spin-dependent scattering. Even slightly below $T_c$ a pure change in the magnetization magnitude occurs which causes the enhancement of damping. Also our experimental results reveal that the precession amplitude and damping have been subjected to a sudden change for F > 30 mJ/cm$^2$. Energy density deposited in the probed volume is proportional to pump fluence. For higher fluence, the temperature dependence of the electronic specific heat plays major role. The increase in the electronic specific heat value with temperature may lead to longer thermal-relaxation time. We infer that relative balance between the energy deposited into the lattice and electron system is also different for higher fluence regime compared to that in the lower fluence regime. Thus, the system temperature remains well above Curie temperature for F > 30 mJ/cm$^2$, during the onset of precession for t $\geq$ 10 ps. This may open up additional energy dissipation channel for the magnetization relaxation process over nanoseconds time scale. Sometimes within very short time scale the spin temperature can go beyond the Curie temperature leading towards formation of paramagnetic state but that is a highly non-equilibrium case [45]. However we believe that in our experiment, even for the high fluence limit and in local thermal equilibrium the ferromagnetic to paramagnetic transition is not observed. Repetitive measurements established the reversibility of the damping parameter and bias-magnetic-field dependence of precessional frequency confirms ferromagnetic nature of the sample.

**C. Frequency modulation and temporal chirping**

Pump fluence also eventually modulates the precessional frequency by introducing temporal chirping in the uniform precession. After immediate arrival of pump pulse, due to enhancement of the surface temperature, the net magnetization is reduced in picosecond time scale which results in chirping of the precessional oscillation. The initial frequency ($f_i$) is reduced with respect to its intrinsic value at a constant field. But when the probed volume cools with time, the spins try to retain their original precessional frequency. Thus, within a fixed time window, the average frequency ($f_{FFT}$) also undergoes slight modification. In the high fluence regime, significant red shift is observed in both $f_{FFT}$ and $f_i$. For $H$ = 2.4 and 1.8 kOe, modulation of frequency is found to be 0.020 GHz.cm$^2$/mJ for $f_{FFT}$ and 0.028 GHz.cm$^2$/mJ for $f_i$, from the slope of linear fit (as shown in Fig. 4(a)). The $f_{FFT}$ is reduced by 7.2% of the extrapolated value at zero pump fluence for both the fields.

On the other hand, $f_i$ is decreased by 8.7% of its zero pump value for the highest pump fluence. The temporal chirp parameter, $b$ shows giant enhancement within the experimental fluence range (Fig. 4(b)). For $H$ = 2.4 kOe, $b$ has increased up to ten times (from 0.03 GHz/ns to 0.33 GHz/ns) in this fluence limit which implies an increase in frequency of 0.66 GHz. Within our experimental scan window (2 ns), the maximum frequency shift is found to be 4.5% for $F$ = 55 mJ/cm$^2$. For another bias field ($H$ = 1.8 kOe), the enhancement of chirp parameter follows the similar trend. This ultrafast modulation is attributed to the thermal effect on the local magnetic properties within the probed volume and is inferred to be reversible [37]. We have also plotted the variation of $b$ with applied bias field for four different pump fluencies. It seems to be almost constant for all the field values in moderate fluence regime (as shown in Fig. 4 (c)). But for $F$ = 40 mJ/cm$^2$, data points are relatively scattered and large errors have been considered to take care of those fluctuations.

## IV. CONCLUSION

In essence, fluence dependent study of ultrafast magnetization dynamics in $Ni_{80}Fe_{20}$ thin film reveals very weak correlation between ultrafast demagnetization time and Gilbert damping within our experimental fluence range. We have reported large enhancement of damping with pump fluence. From the bias field as well as pump fluence dependence of experimentally obtained dynamical parameters we have excluded all the possible extrinsic contributions and observed a pump-induced modulation of intrinsic Gilbert damping. Also, from repetitive measurements with different pump irradiation we have shown that the pump-induced changes are reversible in nature. Enhancement of the system temperature to Curie temperature ratio is believed to be responsible for increment in remagnetization times and damping. The temporal chirp parameter has been found to be increased by up to ten times within the experimental fluence range, while the frequency experiences a significant red shift. From application point of view, as increasing demand for faster and efficient magnetic memory devices, has led the scientific community in the extensive research field of ultrafast magnetization dynamics, our results will further enlighten the understanding of modulation of magnetization dynamics in ferromagnetic systems in presence of higher pump fluence. Usually low damping materials are preferred because it is easier to switch their magnetization in expense of smaller energy, lower write current in STT-MRAM devices and longer propagation length of spin waves im magnonic devices. On the other hand, higher damping is also required to stop the post switching ringing of the signal. The results also have important implications on the emergent field of all-optical helicity dependent switching [46-48]. In this context, the transient modulation of Gilbert damping and other dynamical parameters in ferromagnetic materials is of fundamental interest for characterizing and controlling ultrafast responses in magnetic structures.

**Acknowledgements:** We gratefully acknowledge the financial support from S. N. Bose National Centre for Basic Sciences (grant no.: SNB/AB/12-13/96 and SNB/AB/18-19/211) and Department of Science and Technology (DST), Government of India (grant no.: SR/NM/NS-09/2011). We also gratefully acknowledge the technical assistance of Dr. Jaivardhan Sinha and Mr. Samiran Choudhury for preparation of the sample. SM acknowledges DST for INSPIRE fellowship.

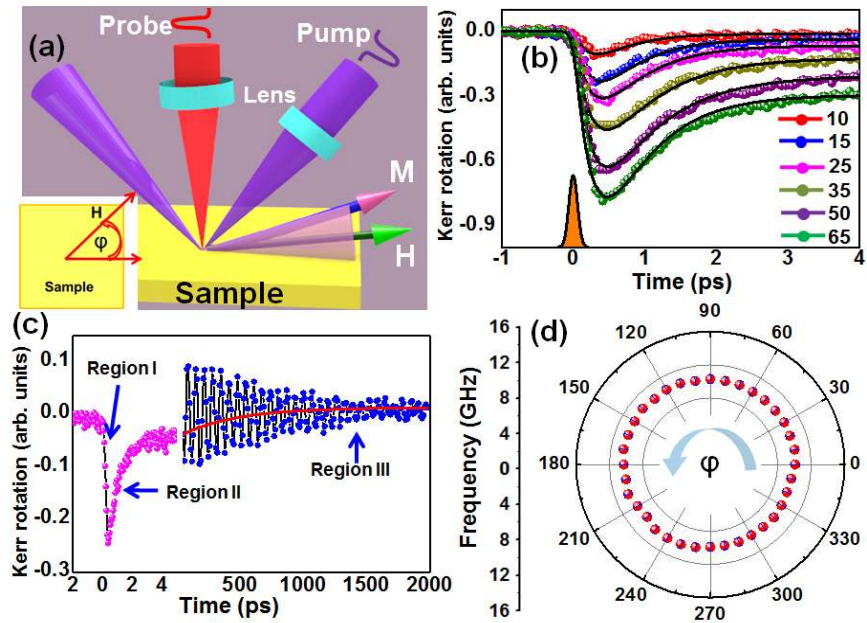

**Figure 1**: (a) Schematic of experimental geometry. In the inset, φ is shown as in-plane rotational angle of H, (b) pump fluence dependence of ultrafast demagnetization; Solid lines are fitting lines. Pump fluences (*F*) having unit of mJ/cm$^2$ are mentioned in numerical figure. The Gaussian envelope of laser pulse is presented to describe the convolution. (c) Representative time resolved Kerr rotation data with three distinguished temporal regions for *F* = 25 mJ/cm$^2$. (d) Angular variation of precessional frequency at *H* = 1.1 kOe for 20-nm-thick Py film. φ is presented in degree.

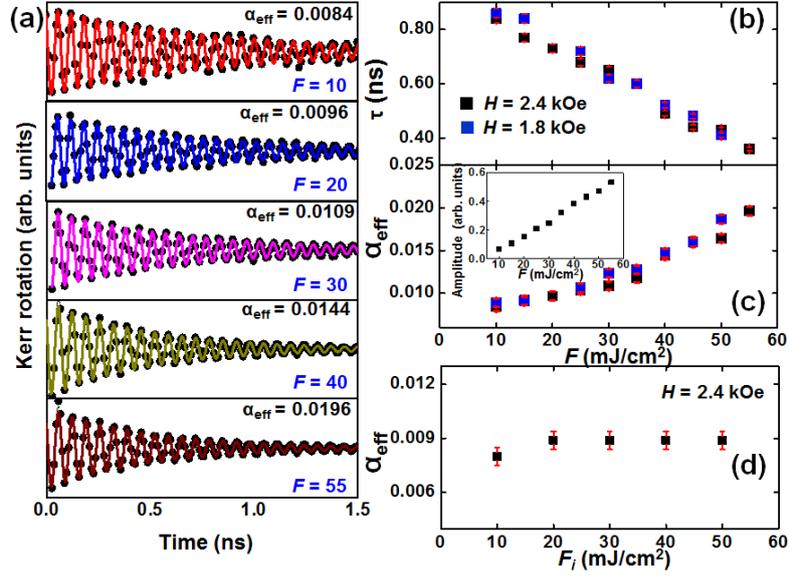

**Figure 2**: (a) Background subtracted time-resolved Kerr rotation data for different pump fluences at $H$ = 2.4 kOe. $F$ having unit of mJ/cm$^2$ is mentioned in numerical figure. Solid lines are fitting lines. Pump fluence dependence of (b) relaxation time ($\tau$) and (c) effective damping ($\alpha_{eff}$). Black and blue symbols represent the variation of these parameters at two different field values, $H$ = 2.4 and 1.8 kOe, respectively. Amplitude of precession is also plotted with pump fluence for $H$ = 2.4 kOe, (d) Variation of effective damping with irradiation fluence ($F_i$) at $H$ = 2.4 kOe. In order to check the possible damage in the sample as high fluence values the pump fluence was taken up to the targeted value of $F_i$ for several minutes followed by reduction of the pump fluence to a constant value of 10 mJ/cm$^2$ and the pump-probe measurement was performed. The damping coefficient is found to be unaffected by the irradiation fluence as shown in (d).

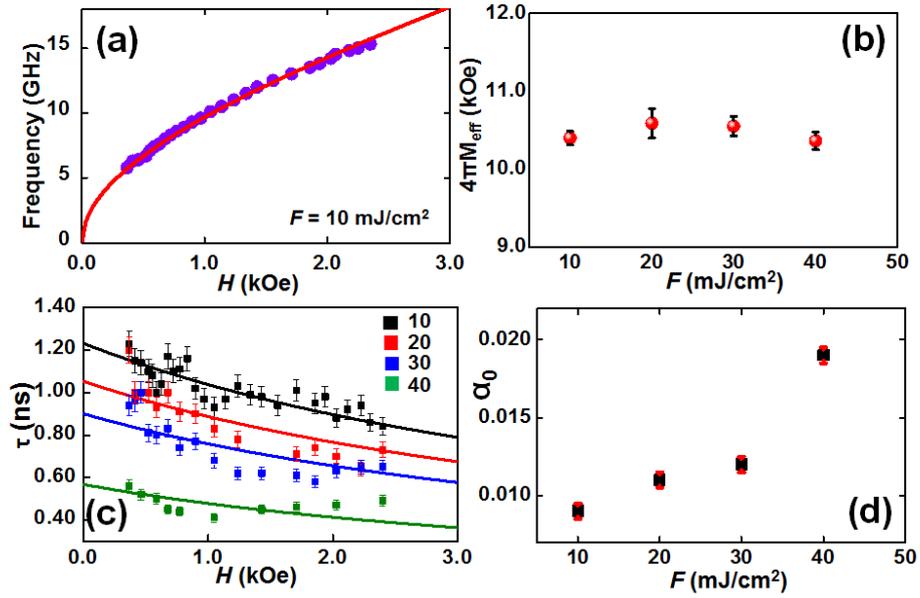

**Figure 3**: (a) Bias field dependence of precessional frequency for $F$ = 10 mJ/cm$^2$. The red solid line indicates the Kittel fit. (b) Pump fluence dependence of effective magnetization ($M_{eff}$) of the probed volume. (c) Bias field dependence of relaxation time ($\tau$) for four different fluences. $F$ having unit of mJ/cm$^2$ is mentioned in numerical figures. Solid lines are the fitted data. (d) Variation of intrinsic Gilbert damping ($\alpha_0$) with pump fluence.

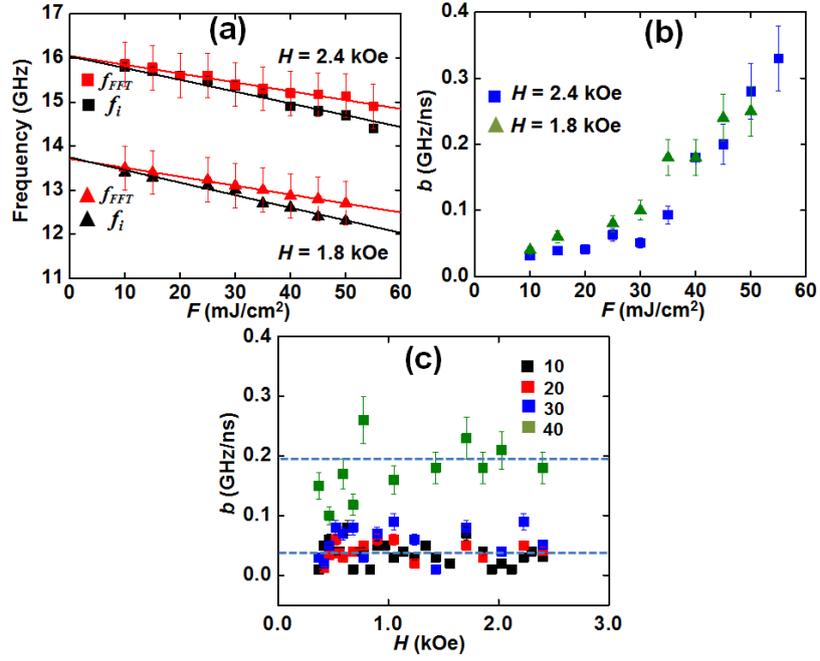

**Figure 4**: (a) Pump-fluence dependence of precessional frequencies for $H$ = 2.4 and 1.8 kOe. Red and black symbols represent the variation of average frequency ($f_{FFT}$) and initial frequency ($f_i$) respectively. (b) Variation of temporal chirp parameter '*b*' with pump fluence for two different magnetic field values. (c) Variation of temporal chirp parameter with bias field for four different pump fluences. *F* having unit of mJ/cm$^2$ is mentioned in numerical figure. Dotted lines are guide to eye.